\documentstyle[11pt]{article}
\begin{document}
\vspace*{-2cm}
\noindent
\hspace*{11cm}
UG--FT--63/96 \\
\hspace*{11cm}
hep--ph/9607311 \\
\hspace*{11cm}
July 1996 \\
\begin{center}
\begin{large}
{\bf Invariant analysis of CP violation \\}
\end{large}
~\\
F. del Aguila, J. A. Aguilar--Saavedra \\
{\it Departamento de F\'{\i}sica Te\'{o}rica y del Cosmos \\
Universidad de Granada \\
18071 Granada, Spain} \\
~\\
M. Zra{\l}ek \\
{\it Field Theory and Particle Physics Dept. \\
University of Silesia \\
40-007 Katowice, Poland}
\end{center}
\begin{abstract}
The invariant formulation of CP violation involves the generation of sets of
invariant constraints for CP conservation, the manipulation of their expressions
and the identification of complete and minimal subsets of such constraints.
In this paper we present a collection of
subroutines to deal with these three tasks in a fast, reliable and systematic
way, with examples for the leptonic sector.
\end{abstract}
\hspace{0.8cm}

\noindent
\hspace{0.8cm}
\section{Introduction}
The observed CP violation in the ${\mathrm K}^0-\bar {\mathrm K}^0$ system 
\cite{papiro1} is related
to the presence of complex phases in the mixing
matrix describing the quark gauge couplings in the mass-eigenstate basis in
the Standard Model (SM) \cite{papiro2}. However, 
not all the phases in the mixing matrices are CP violating. Some of them can be
eliminated redefining the fermion phases. In the SM with three quark
generations,
the 6 phases of the $3 \times 3$ unitary mixing matrix $V$ reduce to 1 
after an appropriate
field redefinition. For this case it has been identified a quantity invariant
under weak quark basis transformations whose vanishing characterizes CP
conservation \cite{papiro3}:
\begin{eqnarray}
\det\; [M_u M_u^\dagger,M_d M_d^\dagger]& = &
-2 i (m_t^2-m_c^2)(m_t^2-m_u^2) (m_c^2-m_u^2)  
 (m_b^2-m_s^2) \nonumber \\ & \times & (m_b^2-m_d^2) (m_s^2-m_d^2)\,
 {\mathrm Im\;}
 (V_{ud} V_{cd}^* V_{cs} V_{us}^{*})=0,
\label{ec:2}
\end{eqnarray}
where $m_i$ is the mass of the quark $i$ and $V_{ij}$ is the $ij$ entry of the
Cabibbo-Kobayashi-Maskawa (CKM) matrix \cite{papiro2}. This invariant 
formulation of CP
violation makes apparent the necessary and sufficient conditions for CP
conservation, giving the size of CP violation. It also allows to decide if CP
is conserved in any weak basis, motivating model building and helping to
understand eventually the origin of CP violation if a definite model (weak
basis) if physically distinguished.

In general, {\em i.e.} in extended models with extra generations and/or
vector-like quarks and/or right-handed (Majorana) neutrinos, it is difficult to
find such a minimal set of necessary and sufficient invariant conditions. In
this paper we present a collection of subroutines for {\em Mathematica}
\cite{papiro3b} which
carries out this task in a fast and systematic way. They are used ({\em i\/}) to 
calculate products of mass (sub)matrices giving rise to expressions invariant 
under weak fermion basis transformations, ({\em ii\/}) to do the symbolic 
manipulations and to solve the constraints resulting from equating these 
invariants to zero, and ({\em iii\/}) to verify that the set of necessary 
conditions for CP conservation generated in this way is sufficient and minimal
(this is to say that this set does not contain any smaller subset of sufficient
conditions). Although these subroutines are written for the leptons including
Majorana neutrinos, they can be easily adapted to analyze CP violation in the
quark sector \cite{papiro4,papiro5}. They can be also generalized to study
extended gauge theories as left-right models, and/or (supersymmetric) scalar
gauge couplings. On the other hand,
 the computer calculations use to be more reliable 
than the manual ones. For example, we have revised the two more complicated
cases studied in Ref. \cite{papiro6}. In both cases we find CP violating
solutions which were overlooked there. In Section 2 we comment on the main
theoretical features and in Section 3 we describe the subroutines
and apply them to these two cases.

\section{Theory and general features}
Let us assume the SM with $n_L$ left-handed and $n_R$ right-handed neutrinos
and define
\begin{equation}
\psi_L=\left ( \begin{array}{c}
\nu_L \\ (\nu^c)_L
\end{array} \right )~,~~~
\psi_R=\left ( \begin{array}{c}
(\nu^c)_R \\ \nu_R
\end{array} \right )~,
\label{ec:3}
\end{equation}
where $\nu^c=i \gamma^2 \nu^*$ and $\psi_{L,R}$ are $n_L+n_R$ component vectors
describing the $n_L$ left-handed and $n_R$ right-handed neutrinos. Then the mass
term reads
\begin{equation}
-{\mathcal L}_{mass}=\bar l_L M_l l_R+ \frac{1}{2} \bar \psi_L M_\nu \psi_R +
{\mathrm h.c.,}
\label{ec:4}
\end{equation}
where $l_{L,R}$ are $n_L$ component vectors describing the $n_L$ charged leptons
and $M_l$ an arbitrary $n_L \times n_L$  complex matrix, whereas
\begin{equation}
M_{\nu} = 
{\begin{array}{c} 
n_L \{ \\ n_R \{
\end{array}}
{ \overbrace{M_L}^{n_L} \ \overbrace{M_D}^{n_R} \choose M^T_D \ M_R }
\label{ec:4b}
\end{equation}
is an arbitrary $(n_L+n_R) \times  (n_L+n_R)$ complex symmetric matrix. This
Lagrangian is invariant under a CP transformation leaving the gauge $\mathrm
SU(2)_L \times U(1)_Y$ interactions unchanged \cite{papiro8}
\begin{equation}
l_L\rightarrow U_L C l_L^* ~,~~ l_R\rightarrow U_R^l C l_R^*~,~~
\nu_L\rightarrow U_L C \nu_L^* ~,~~ \nu_R\rightarrow U_R^\nu C \nu_R^* ~,
\label{ec:5}
\end{equation}
where $C$ is the Dirac charge conjugation matrix and
$U_L$, $U_R^l$ ($U_R^\nu$) are $n_L \times n_L$ ($n_R \times n_R$) unitary 
matrices, if
\begin{equation}
U_L^\dagger M_l U_R^l  =  M_l^* ~,~
U_L^\dagger M_L U_L^*  =  M_L^* ~,~
U_L^\dagger M_D U_R^\nu  =  M_D^* ~,~
{U_R^\nu}^T M_R U_R^\nu  =  M_R^*~.
\label{ec:6}
\end{equation}
Eqs. (\ref{ec:6}) are necessary and sufficient conditions for CP conservation.
From these equations, one can write necessary conditions for CP conservation  
which do not require the knowledge of the unitary
matrices involved in the definition of the CP transformation. Thus, the products
of mass matrices $M_l$, $M_L$, $M_D$, $M_R$ are classified in three 
classes $G_1 \equiv G_L$, $G_2 \equiv G_R^l$, $G_3 \equiv G_R^\nu$, depending 
under which unitary matrix 
($U_L$, $U_R^l$, $U_R^\nu$, respectively) transform,
\begin{eqnarray}
G_1 & = & \{  A_{L1} \equiv M_l M_l^\dagger, A_{L2} \equiv M_L M_L^\dagger, 
A_{L3} \equiv M_D 
M_D^\dagger, A_{Li} A_{Lj}, M_L M_l^* M_l^T M_L^\dagger, \nonumber \\
& &  M_L M_D^* M_D^T M_L^\dagger, 
M_L M_D^* M_R M_D^\dagger, M_D M_R^\dagger M_D^T M_L^\dagger,
M_D M_R^\dagger M_R M_D^\dagger, \dots \}, \nonumber \\
G_2 & = & \{ A_l \equiv M_l^\dagger M_l,A_l^2, M_l^\dagger M_L M_L^\dagger 
M_l, M_l^\dagger M_D M_D^\dagger M_l, \dots \}, \nonumber \\
G_3 & = & \{A_{\nu 1} \equiv M_D^\dagger M_D, A_{\nu 2} \equiv M_R^\dagger 
M_R,A_{\nu i} A_{\nu j},M_D^\dagger M_l M_l^\dagger M_D,
M_D^\dagger M_L M_L^\dagger M_D, \nonumber \\
& & M_D^\dagger M_L M_D^* M_R,M_R^\dagger M_D^T M_L^\dagger M_D,
M_R^\dagger M_D^T M_D^* M_R, \dots \}. \label{ec:7}
\end{eqnarray}
The trace and determinant of any of such products or their sums within the same
class do not change under unitary transformations due to their cyclic and
factorization properties, respectively. Then CP conservation (Eq. (\ref{ec:6}))
implies that the trace and the determinant of any linear combination $g$ of 
elements of the same class is real,
\begin{equation}
{\mathrm Im\;tr\;}g = 0 ~,~~~
{\mathrm Im\;}\det g = 0.
\label{ec:8}
\end{equation}
To find which subsets of these conditions are also sufficient has to be worked
out case by case. In practice, we are interested in identifying minimal subsets
with these properties. We proceed as follows.

\subsection{Generation of invariant products with the same transformation
properties}
The elements of $G_i$ in Eq. (\ref{ec:7}) must be generated using the mass
matrices
\begin{equation}
\{ M_l,M_l^T,M_l^*,M_l^\dagger,M_L,M_L^\dagger,M_D,M_D^T,M_D^*,M_D^\dagger,
M_R,M_R^\dagger \}.
\label{ec:9}
\end{equation}
(Note that $M_{L,R}$ are symmetric.) A sequence $S$ of order $n$ is an allowed
product of $n$ of such matrices. A product of two sequences $S_a S_b$ is allowed
if $S_a'=V_a S_a V_a'^\dagger$, $S_b'=V_b S_b V_b'^\dagger$ and $V_a'=V_b$.
Hence the sequences of order 2 are
\begin{eqnarray}
& & \{ M_l M_l^\dagger,  M_l^T M_l^*, M_l^T M_L^\dagger, M_l^T M_D^*, 
M_l^* M_l^T, M_l^\dagger M_l, M_l^\dagger M_L, M_l^\dagger M_D, \nonumber \\
& & M_L M_l^*, M_L M_L^\dagger, M_L M_D^*, M_L^\dagger M_l, M_L^\dagger M_L,
 M_L^\dagger M_D, M_D M_D^\dagger, M_D M_R^\dagger, \nonumber \\
& & M_D^T M_l^*, M_D^T M_L^\dagger, M_D^T M_D^*, M_D^* M_D^T, M_D^* M_R,
M_D^\dagger M_l, M_D^\dagger M_L, M_D^\dagger M_D, \nonumber \\
& & M_R M_D^\dagger, M_R M_R^\dagger, M_R^\dagger M_D^T, M_R^\dagger M_R  \}.
\label{ec:10}
\end{eqnarray}

A sequence $S$, with $S'=V S V'^\dagger$ is an element of $G_{1,2,3}$ in Eq. 
(\ref{ec:7}) if $V=V'=U_L, U_R^l, U_R^\nu$, respectively. Then Eq. (\ref{ec:6})
implies that the order of the elements of $G_i$ is even and that the sequences
in $G_i$ can be constructed with the biproducts in Eq. (\ref{ec:10}). These
elements are generated order by order. The elements of order $n$ are the
allowed products of the sequences of order $n-2$ which begin with $M_l$, $M_L$
or $M_D$ for $G_1$, $M_l^\dagger$ for $G_2$ and $M_D^\dagger$ or $M_R^\dagger$
for $G_3$ (see Eq. (\ref{ec:6})), times the biproducts in Eq. (\ref{ec:10})
which end with $M_l^\dagger$, $M_L^\dagger$ or $M_D^\dagger$ for $G_1$, $M_l$ 
for $G_2$ and $M_D$ or $M_R$ for $G_3$.

\subsection{Solution of invariant constraints}
Once the elements of $G_i$ are generated, we have to solve the nontrivial
conditions in Eq. (\ref{ec:8}) up to a given order. This means to find the
relations among the parameters fixing $M_{l,L,D,R}$ implied by these
conditions. The order is increased until there is no CP violating solution.
In the examples discussed later we need to consider conditions involving
sequences of order 4, 6 and 8. These constraints are easier to solve if the mass
matrices are conveniently parametrized. Under a change of weak basis,
\begin{eqnarray}
M_l' & = & W_L M_l {W_R^l}^\dagger ~~,~~
M_L' = W_L M_L W_L^T, \nonumber \\
M_D' & = & W_L M_D {W_R^\nu}^\dagger ~ ,~~ 
M_R' = {W_R^\nu}^\dagger M_R {W_R^\nu}^\dagger~.
\label{ec:11}
\end{eqnarray}
Hence, as the constraints do not depend on the choice of weak basis, we can
choose the unitary matrices $W_L$, $W_R^{l,\nu}$ appropriately and assume $M_l$
and $M_R$ diagonal with nonnegative real elements, whereas $M_L$ is
complex and symmetric and $M_D$ complex and arbitrary. The number and
difficulty of the equations grow with the order of the sequences. Generically
{\em Mathematica} uses too much memory and time to solve simultaneously more
than 5 conditions. In this case one solves the equations of lowest order first
and inserts the solutions in the remaining equations, which can then be solved
more easily.

\subsection{Minimal subset of invariant constraints}
The last subroutine verifies if a solution of a set of
conditions in Eq. (\ref{ec:8}) is CP conserving. A solution $s$ is a set of
relations among the parameters fixing the mass matrices $M_{l,L,D,R}$. Let 
$M_{l,L,D,R}^{(s)}$ be the mass matrices satisfying these relations. We generate
random numbers for their independent parameters denoting the corresponding mass
matrices $\tilde M_{l,L,D,R}^{(s)}$.
These matrices conserve CP if it exists
a unitary matrix $\tilde U^{(s)}$ preserving $\tilde M_l^{(s)}$ and such that 
\begin{equation}
\tilde U^{(s)} \tilde M_\nu^{(s)} \tilde U^{(s)T}=\tilde M_\nu^{(s)*}.
\label{ec:14}
\end{equation}
If this matrix can be constructed, we say that $s$ conserves CP. It is highly 
improbable that there is a relation among the random parameters in 
$\tilde M_{l,L,D,R}^{(s)}$ such that the matrices $M_{l,L,D,R}^{(s)}$ violate CP
but $\tilde M_{l,L,D,R}^{(s)}$ do not. On the other hand, if 
Eq. (\ref{ec:14}) is not satisfied, the solution $s$ is said to violate CP.

Let us drop the superscripts from now on. 
The existence of $U$ is determined in two steps. Step 1 
guarantees that
if $U$ exists, it is diagonal, and step 2 checks if there is a diagonal
unitary matrix satisfying Eq. (\ref{ec:14}). In the convenient basis and for a
given solution $s$, $U$ is block diagonal with diagonal submatrices of
dimension the multiplicity $m_i$ and $\mu_i$ of the eigenvalues of $M_l$
and $M_R$ respectively,
\begin{equation}
U= \left( \;
\begin{array}{cccccc}
\cline{1-1}
\multicolumn{1}{|c|}{(m_1 \times m_1)} \\
\cline{1-2}
& \multicolumn{1}{|c|}{(m_2 \times m_2)} \\
\cline{2-2}
& & ~\ddots~ \\
\cline{4-4}
& & & \multicolumn{1}{|c|}{(\mu_1 \times \mu_1)} \\
\cline{4-5}
& & & & \multicolumn{1}{|c|}{(\mu_2 \times \mu_2)} \\
\cline{5-5}
& & & & & ~\ddots~
\end{array}\; \right),
\end{equation}
with $\sum m_i=n_L$, $\sum \mu_i=n_R$.

If there is no degeneracy $U$ is diagonal and we go to step 2: Eq. (\ref{ec:14})
is fulfilled if $\forall i>j$,
\begin{equation}
\Delta_{ij} \equiv \arg M_{ii}-\arg M_{ij}+\arg M_{jj}-\arg M_{ji}=0,\pi.
\label{ec:15}
\end{equation}
If any $M_{ij}=0$, $\Delta_{ij}$ is zero for $\arg M_{ij}$ 
($=\arg M_{ji}$) is conventionally defined by Eqs. (\ref{ec:15}). On the other 
hand, if some diagonal elements
$M_{i_1 i_1}=\cdots=M_{i_k i_k}=0 $, we consider
$\arg M_{i_1 i_1},\dots,\arg M_{i_k i_k}$ arbitrary but fixed. We regard 
Eqs. (\ref{ec:15}) as a system of equations in the variables $\delta_i \equiv
\arg M_{ii}$. Then CP is conserved if and only if the full system of Eqs. 
({\ref{ec:15}) is compatible.

If there is degeneracy we can fix the weak basis to ensure that $U$ is
diagonal and then go to step 2. This involves the most delicate casuistry,
and it has to be done and programmed for each dimension $n_L$, $n_R$. For small
dimensions it can be easier to do it by hand. Let us discuss the
simplest case: $n_L=1$, $n_R=2$, and the $2 \times 2$ matrix $M_R$
diagonal, real, nonnegative and degenerate. Then $U$ has the general form
\begin{equation}
U=\left( \begin{array}{cc}
e^{-i \delta_L} & 0 \\
0 & U_R^T \end{array} \right).
\end{equation}
\begin{itemize}
\item If $M_R$ is identically zero, $U_R$ is an arbitrary unitary matrix. There
are different ways to fix the weak basis and guarantee that $U_R$ is diagonal.
We choose to diagonalize $M_D^\dagger M_D$.
\begin{itemize}
\item If $M_D^\dagger M_D$ is nondegenerate, we transform $M$ accordingly
and go to step 2.
\item If $M_D^\dagger M_D$ is degenerate, it is identically zero because $M_D$
is a $1 \times 2$ matrix. Then $M_D$ is also identically zero and there is an
extra flavour symmetry, and Eq. (\ref{ec:14}) is satisfied taking $U_R$ equal to
the identity.
\end{itemize}
\item If $M_R$ is nonzero, $U_R$ is a real orthogonal matrix. In this case we
choose to diagonalize $({\mathrm Im}\,M_D)^T ({\mathrm Im}\,M_D)$ or
$({\mathrm Re}\,M_D)^T ({\mathrm Re}\,M_D)$ if the former is degenerate.
\begin{itemize}
\item If one of them is nondegenerate, we transform $M$ accordingly and go to
step 2. (For convenience only, if $M_L=0$ we also require $M_D M_D^T$ to be
real and positive. This fixes $\delta_L$. If $M_D M_D^T=0$ there is a flavour
symmetry.)

\item If both are degenerate, as above they are identically zero. $M_D$ is also
zero and there is again an extra flavour symmetry. Similarly Eq. (\ref{ec:14})
is fulfilled with $U_R$ equal to the identity.
\end{itemize}
\end{itemize}

\section{Examples}
In this section we present a description of the different subroutines in the
package {\tt CPlep}. Some of them require the subroutines {\tt DiagonalizeH}
and {\tt DiagonalizeS} defined in the package {\tt Diagon} \cite{papiro9}.
{\tt DiagonalizeH} is a subroutine which diagonalizes hermitian matrices, 
returning orthonormal eigenvectors even in the case of degenerate 
eigenvalues\footnote{The built-in function {\tt Eigensystem} may give
non-orthogonal eigenvectors}. {\tt DiagonalizeS}
diagonalizes general complex symmetric matrices $M_s$ with a
congruent transformation $M_s \rightarrow U M_s U^T=D$. Let us show how
{\tt CPlep} works in two examples. (Definitions are given in the {\em
Appendix}.)

\subsection{Case $n_L=1$, $n_R=2$} 
The list
\begin{verbatim}
l1={a[10],a[11],a[12],a[13],a[20],a[23],a[30],a[31],a[32],a[33],
a[40],a[43]}
\end{verbatim}
is the computer representation of the mass matrices in Eq. (\ref{ec:9}). We
define an adequate output format for {\tt a[$i$]} so that the expressions can be 
more easily
understood. The rules for the generation of invariants are embodied in the
definition of an
associative function {\tt f}, with an arbitrary number of arguments
{\tt a[$i_1$],$\dots$,\tt a[$i_m$]} ordered as the corresponding matrix product 
{\tt a[$i_1$]$\cdots$a[$i_m$]}. 
{\tt f} is taken to be zero if any of the products {\tt a[$i_1$]a[$i_2$]},
 $\dots$
,{\tt a[$i_{m-1}$]a[$i_m$]} is not allowed (see Eq. (\ref{ec:10})). Then the 
function {\tt inv} checks if
the product {\tt f[a[$i_1$],$\dots$,a[$i_m$]]} transforms with a matrix and its
adjoint, and returns again 
its
argument if it does or 0 if it does not. The lists obtained in this way are
multiplied with {\tt Outer}.
\begin{verbatim}
<<cplep.m;
l2[1]=Union[Flatten[Outer[f,cl1ini,l1]]];
l2[2]=Union[Flatten[Outer[f,cl2ini,l1]]];
l2[3]=Union[Flatten[Outer[f,cl3ini,l1]]];
l3[1]=Union[Flatten[Outer[f,l2[1],l1]]];
l3[2]=Union[Flatten[Outer[f,l2[2],l1]]];
...
l6[3]=Union[Flatten[Outer[f,l5[3],l1]]];

i4[1]=Union[inv[l4[1]]];
i4[2]=Union[inv[l4[2]]];
i4[3]=Union[inv[l4[3]]];
i6[1]=Union[inv[i6[1]]];
i6[2]=Union[inv[i6[2]]];
i6[3]=Union[inv[i6[3]]];
\end{verbatim}
In this example it is necessary to consider only sequences up to order 6. 
In order to obtain the invariant constraints for $n_L=1$, $n_R=2$, we introduce
the convenient parametrization for {\tt Ml}, {\tt ML}, {\tt MD}, {\tt MR}
and define {\tt Mn}, the neutrino mass matrix in Eq. (\ref{ec:4b}). Random
numerical values for the parameters are also generated for later use.
\begin{verbatim}
Ml:={{e1}};ML={{n1}};MD:={{Re[n2]+I Im[n2],Re[n3]+I Im[n3]}};
MR:={ {n4,0},{0,n5} }; Mn=Transpose[Join[Transpose[Join[ML,
Transpose[MD]]], Transpose[Join[MD,MR]]]]; 
ev1={e1->Random[],n1->Random[],Re[n2]->Random[Real,{-1,1}], 
Im[n2]->Random[Real,{-1,1}],Re[n3]->Random[Real,{-1,1}], 
Im[n3]->Random[Real,{-1,1}],n4->Random[],n5->Random[]} 
\end{verbatim}
Now to identify those constraints in Eq. (\ref{ec:8}) which are not identically
zero, we define the functions
{\tt Looktrace} and {\tt Lookdet}. Their first
argument {\tt l} is a list of sequences, whereas the other 
arguments are the mass matrices with random entries. These are required to
check if the constraint is identically zero.
{\tt Looktrace} looks for the elements
$g$ of {\tt l} which satisfy ${\mathrm Im\;tr}\;g \neq 0$. {\tt Lookdet} first
finds all non-hermitian combinations $g-g'$  and then,
returns those with ${\mathrm Im}\;\det(g-g')\neq 0$. Finally {\tt Newecs}
calculates the non-trivial constraints returned by {\tt Looktrace} and 
{\tt Lookdet} as a function of the convenient parameters (see the {\em
Appendix} for a detailed description of its arguments). Thus
\begin{verbatim}
Looktrace[i4[1],Ml/.ev1,ML/.ev1,MD/.ev1,MR/.ev1];
Newecs[{},%,Ml,ML,MD,MR,{e1,n1,n2,n3,n4,n5},{n2,n3}]
ecs={%[[3,2]]};
\end{verbatim}
returns
\begin{equation}
{\mathrm Im\;tr}\;(M_D M_R^\dagger M_D^T M_L^\dagger)=2 n_1(n_4\, {\mathrm Im}\,
n_2\,{\mathrm Re}\,n_2+n_5\, {\mathrm Im}\,n_3\, {\mathrm Re}\,n_3).
\label{ec:17}
\end{equation}
The traces of the other elements of {\tt i4[1]}, {\tt i4[2]} and 
{\tt i4[3]} are proportional to the trace in Eq. (\ref{ec:17}). Similarly,
\begin{verbatim}
Looktrace[i6[1],Ml/.ev1,ML/.ev1,MD/.ev1,MR/.ev1];
Newecs[ecs,%,Ml,ML,MD,MR,{e1,n1,n2,n3,n4,n5},{n2,n3}]
AppendTo[ecs,%[[3,2]]];
\end{verbatim}
gives
\begin{equation}
{\mathrm Im\;tr}\;(M_D M_R^\dagger M_R M_R^\dagger M_D^T M_L^\dagger) = 
2 n_1(n_4^3\, {\mathrm Im}\,n_2\,{\mathrm Re}\,n_2+n_5^3\, {\mathrm Im}\,n_3\,
{\mathrm Re}\,n_3).
\label{ec:18}
\end{equation}
The traces of the remaining sequences in {\tt i6[1]}, {\tt i6[2]} and 
{\tt i6[3]} do not give new equations. Finally
\begin{verbatim}
Lookdet[i4[3],Ml/.ev1,ML/.ev1,MD/.ev1,MR/.ev1];
Newecs[ecs,%,Ml,ML,MD,MR,{e1,n1,n2,n3,n4,n5},{n2,n3},det->True]
AppendTo[ecs,%[[3,3]]];
\end{verbatim}
returns
\begin{eqnarray}
{\mathrm Im}\;\det (M_D^\dagger M_D M_R^\dagger M_R-M_R^\dagger M_D^T M_D^* 
M_R)& = & 2 n_4\,n_5\,(n_4-n_5)\,(n_4+n_5) \nonumber \\
& \times &({\mathrm Im}\,n_2\,{\mathrm Re}\,n_3-{\mathrm Im}\,n_3\,
{\mathrm Re}\,n_2) \nonumber \\
& \times &({\mathrm Im}\,n_2\,{\mathrm Im}\,n_3+{\mathrm Re}\,n_2\,
{\mathrm Re}\,n_3).
\label{ec:19}
\end{eqnarray}
When we consider that the set of constraints may be complete, we use 
{\tt Reduce} to find all the solutions. In this case we try with Eqs.
(\ref{ec:17}--\ref{ec:19}) equal to zero. {\tt Reduce} gives a lot of redundant
solutions, often
repeated, for instance {\tt n1==0 \&\& n4==0} and {\tt n1==0 \&\& n4==0 \&\& 
n5==0}, and some inconsistent solutions, like {\tt Re[n3]==I Re[n2]}. The 
function {\tt Eliminatesols} gets rid of the redundant
solutions; whereas the inconsistent solutions are eliminated by inspection.
\begin{verbatim}
sol0=Reduce[ecs=={0,0,0}];
<<Algebra/ReIm.m;
e1/: Positive[e1] = True;n1/: Positive[n1] = True;
n4/: Positive[n4] = True;n5/: Positive[n5] = True;
sol1=Eliminatesols[sol0/._!=0->True];
sol2=sol1/.{(Re[_]==Complex[_,_] x_/;Im[x]==0)->False,
(Im[_]==Complex[_,_] x_/;Im[x]==0)->False};
\end{verbatim}
In this case, {\tt Reduce} gives us 198 solutions ({\tt sol0}). Many of them
are redundant and we need to keep only 32 ({\tt sol1}). 6 of them are
inconsistent and are eliminated in {\tt sol2}.
To verify whether the former set is complete, we use the subroutines {\tt Red12}
 and
{\tt LookCP}. {\tt Red12} writes the mass matrix {\tt Mn} in the basis in which
$U$ is diagonal, if it exists. {\tt LookCP} then
returns {\tt True} or {\tt False} depending on whether $U$ exists.

Although the whole process has been described in the
previous section, we want to point out a little trick used in
{\tt LookCP}. When trying to solve the system of equations $\Delta_{ij}=0,\pi$
in Eqs. (\ref{ec:15}), one could consider instead to solve the system of
equations $\sin \Delta_{ij}=0$. However, {\em Mathematica} does not give
correct results in this case. For this reason we look for a subset of 
{\em linearly independent} equations in Eqs. (\ref{ec:15}), and solve these
equations equated to zero (it can be done always if the equations are
independent, because we can always redefine the variables $\delta_i = \arg
M_{ii}$ conveniently).      
Then the solution is substituted in the complete set of equations, checking 
whether they are equal to $0,\pi$.

We check if any of the 26 solutions in {\tt sol2} violates CP using a
loop. It is initialized with
\begin{verbatim}
m[number_]:=Mn/.ToRules[sol2[[number]]]/.ev1;
i=0;
\end{verbatim}
and for each solution we run
\begin{verbatim}
++i;LookCP[Red12[m[i]]]
\end{verbatim}
The result is that none of the solutions in {\tt sol2} violates CP. So the set
of invariant constraints is complete. We could try to take away any of the 
equations in {\tt ecs} and repeat the same process, but we would find CP 
violating solutions. So the set is also minimal.
\subsection{Case $n_L=2$, $n_R=1$}
This case is more difficult to solve because more
invariant constraints are necessary. As we pointed out in Section 2, the 
computer
memory required to solve all the equations simultaneously is too big, so
we have to use a different approach. First we solve the condition
\begin{equation}
{\mathrm Im}\;\det (M_L M_L^\dagger M_l M_l^\dagger-M_L M_l^* M_l^T
M_L^\dagger) = 0.
\label{ec:16}
\end{equation}
Then for each solution of Eq. (\ref{ec:16}) we proceed as in the former example
but restricting the general form of the mass matrices to fulfill this
particular solution. For this case we have found that
\begin{eqnarray}
{\mathrm Im\;tr}\;(M_D^\dagger M_L M_D^* M_R) & = & 0, \nonumber \\
{\mathrm Im\;tr}\;(M_R^\dagger M_D^T M_L^\dagger M_L M_L^\dagger M_D)
& = & 0, \nonumber \\
{\mathrm Im\;tr}\; (M_l M_l^\dagger M_L M_l^* M_l^T M_D^* M_R M_D^\dagger)
& = & 0, \nonumber \\
{\mathrm Im}\;\det (M_L M_L^\dagger M_l M_l^\dagger-M_L M_l^* M_l^T
M_L^\dagger) & = & 0, \nonumber \\
{\mathrm Im}\;\det (M_D M_D^\dagger M_L M_L^\dagger-M_L M_D^* M_D^T 
M_L^\dagger ) & = & 0, \nonumber \\
{\mathrm Im}\;\det (M_D M_D^\dagger M_l M_l^\dagger-M_L M_D^* M_D^T 
M_L^\dagger ) & = & 0, \nonumber \\
{\mathrm Im}\;\det (M_D M_D^\dagger M_l M_l^\dagger-M_l M_l^\dagger
M_L M_L^\dagger) & = & 0, \nonumber \\
{\mathrm Im}\;\det (M_D M_D^\dagger M_l M_l^\dagger-M_L M_l^* M_l^T
M_L^\dagger) & = & 0, \nonumber \\
{\mathrm Im}\;\det (M_L M_D^* M_R M_D^\dagger-M_l M_l^\dagger M_l 
M_l^\dagger) & = & 0
\end{eqnarray}
form a complete set of constraints. This set is also minimal by construction,
as long as Eq. (\ref{ec:16}) is included.

\vspace{1cm}
\noindent
{\Large \bf Acknowledgements}

\vspace{0.4cm}
\noindent
We thank G. Branco for discussions. This work was
partially supported by CICYT under contract AEN94-0936, by the Junta
de Andaluc\'{\i}a and by the European Union under contract
CHRX-CT92-0004.
\newpage
\appendix
\section{Appendix}
\begin{verbatim}
(* 
This package is available by anonymous ftp at deneb.ugr.es 
in directory pub/packages
*)

BeginPackage["CPlep`","Diagon`"]

(*
This package needs the functions DiagonalizeS, DiagonalizeH
defined in the package Diagon, available by anonymous ftp at 
deneb.ugr.es 
   
DiagonalizeS[m] gives a list {v,u} where m is a square matrix,
 v is a row vector and u is a matrix fulfilling 
 u.m.Transpose[u]=DiagonalMatrix[v]

DiagonalizeH[m] is a modification of Eigensystem which gives a
 list {v,u}, with the eigenvalues and eigenvectors of the 
 hermitian matrix m, such that 
 u.m.Transpose[Conjugate[u]]=DiagonalMatrix[v]. The list of 
 eigenvalues v is ordered by increasing absolute value
*)
    
cl1ini::usage =
 "cl1ini is a list with the possible first elements of a product in
 the class G1"
cl2ini::usage =
 "cl2ini is a list with the possible first elements of a product in
 the class G2"
cl3ini::usage =
 "cl3ini is a list with the possible first elements of a product in
 the class G3"
cl1fin::usage =
 "cl1fin is a list with the possible last elements of a product in
 the class G1"
cl2fin::usage =
 "cl2fin is a list with the possible last elements of a product in
 the class G2"
cl3fin::usage =
 "cl3fin is a list with the possible last elements of a product in
 the class G3"
l1::usage =
 "l1 is a list with all the possible elements of a product"
f::usage =
 "f[expr] is 0 if expr is not an allowed product"
inv::usage =
 "inv[expr] returns expr if it is an invariant product, or 0 if not"
Looktrace::usage =
 "Looktrace[lf,Ml,ML,MD,MR] returns the elements in a list lf of
 f-products with trace nonzero when evaluated with the numerical
 matrices Ml,ML,MD,MR. The output is in the form of Dot products"
Lookdet::usage =
 "Lookdet[lf,Ml,ML,MD,MR] calculates all possible non-hermitian
 combinations x-y of elements of a list lf of f-products and then 
 returns those with nonzero determinant when evaluated with the 
 numerical matrices Ml,ML,MD,MR. The output is in the form of Dot
 products"
Newecs::usage =
 "Newecs[oldecs,ld,Ml,ML,MD,MR,vars,comp] looks for equations not
 present in oldecs, generated with a list ld of Dot products and the
 matrices Ml,ML,MD,MR with variables vars, of which comp are complex.
 It returns a list containing four lists: the list oldecs, possibly
 improved with simpler equations, the list of elements of ld that
 improve oldecs, a list of new equations not present in oldecs and a
 list of elements of ld that generate the new equations"
det::usage =
 "det is an option for Newecs with default value False which select 
 to calculate the trace or the determinant"
Eliminatesols::usage =
 "Eliminatesols[sols] eliminates redundant solutions in sols"
Red12::usage = 
 "Red12[M] reduces a neutrino mass matrix M in the case nL=1,nR=2 by
 changes of basis to a form in which CP is a diagonal matrix if it
 exists"
Red21::usage =
 "Red21[M,Me] reduces a neutrino mass matrix M in the case nL=2,nR=1
 by changes of basis to a form in which CP is a diagonal matrix if it
 exists. Me is the charged lepton mass matrix"
LookCP::usage =
 "LookCP[M] tests if a diagonal unitary matrix U exists such that
 U.M.Transpose[U]=Conjugate[M] and yields True or False" 


Begin["`Private`"]

(* Definitions for internal use only *)

(* No comments needed *)

trace[x_?MatrixQ]:=Sum[x[[i,i]],
{i,Length[x]}]/;Length[x]==Length[Transpose[x]];

Adj[m_]:=Transpose[Conjugate[m]]

(* Enlarge the dimension of a matrix with the identity *)

IncDim1[m_]:=Transpose[Prepend[Transpose[Prepend[m,Table[0,
{Length[m]}]]],Join[{1},Table[0,{Length[m]}]]]]

IncDim[m_,n_]:=Nest[IncDim1,m,n]

(* adjoint of a matrix *)

ad1[a[x_]]:=a[10 Floor[x/10]+3-Mod[x,10]]; 

(* Adjoint of an f-product *)

ad[f[x__]]:=Apply[f,Reverse[Thread[ad1[Apply[ff,f[x]]],ff]]];   
ad[0]=0;

(* Take hermitian elements of a list *)

auto[l_List]:=Block[{lt},(
For[lt={};n=1,n<=Length[l],n++,If[l[[n]]==ad[l[[n]]],
AppendTo[lt,l[[n]]]]];lt)];

(* Take non-hermitian elements of a list *)

autono[l_List]:=Block[{lt},(
For[lt={};n=1,n<=Length[l],n++,If[l[[n]]==ad[l[[n]]],,
AppendTo[lt,l[[n]]],AppendTo[lt,l[[n]]]]];lt)];

(* Imtr and Imdet are used by Newecs and just substitute the
symbolic expressions of the mass matrices and calculate the
imaginary part of the trace/determinant *)

Imtr[expr_,x_,y_,u_,v_,comp_]:=Block[{sust},(
sust={a[10]->x,a[20]->y,a[30]->u,a[40]->v,
a[13]->Adj[x],a[23]->Adj[y],a[33]->Adj[u],a[43]->Adj[v],
a[11]->Transpose[x],a[31]->Transpose[u],
a[12]->Conjugate[x],a[32]->Conjugate[u]};
Factor[ComplexExpand[Im[trace[expr/.sust]],comp] ] )]

Imdet[expr_,x_,y_,u_,v_,comp_]:=Block[{sust},(
sust={a[10]->x,a[20]->y,a[30]->u,a[40]->v,
a[13]->Adj[x],a[23]->Adj[y],a[33]->Adj[u],a[43]->Adj[v],
a[11]->Transpose[x],a[31]->Transpose[u],
a[12]->Conjugate[x],a[32]->Conjugate[u]};
Factor[ComplexExpand[Im[Det[expr/.sust]],comp] ] )]

(* minors calculates the set of invariant phases \Delta_{ij} of a 
matrix. Used by LookCP *)

minor[m_,i_,j_]:=m[[i,i]]+m[[j,j]]-m[[i,j]]-m[[j,i]]

minors[m_]:=Flatten[Table[If[m[[i,j]]==0,0,minor[m,i,j],
minor[m,i,j]],{i,1,Length[m]},{j,i+1,Length[m]}]] 

(* vec and linealindep check if a susbset of invariant phases is
linearly independent. Used by LookCP *)

vec[expr_,d_]:=Table[Coefficient[expr,delta[i]],{i,d}] 

linealindep[exprs_,d_]:=MemberQ[Complement[Flatten[Minors[Thread[
vec[exprs,d]],Length[exprs]]],{0}],_]


(* Output format with standard notation *)

Format[f[x_,y__]]:={x,y};
Format[a[10]]=ColumnForm[{" ","M"," l"},Left,Center];
Format[a[13]]=ColumnForm[{" +","M"," l"},Left,Center];
Format[a[12]]=ColumnForm[{" *","M"," l"},Left,Center];
Format[a[11]]=ColumnForm[{" T","M"," l"},Left,Center];
Format[a[20]]=ColumnForm[{" ","M"," L"},Left,Center];
Format[a[23]]=ColumnForm[{" +","M"," L"},Left,Center];
Format[a[30]]=ColumnForm[{" ","M"," D"},Left,Center];
Format[a[33]]=ColumnForm[{" +","M"," D"},Left,Center];
Format[a[32]]=ColumnForm[{" *","M"," D"},Left,Center];
Format[a[31]]=ColumnForm[{" T","M"," D"},Left,Center];
Format[a[40]]=ColumnForm[{" ","M"," R"},Left,Center];
Format[a[43]]=ColumnForm[{" +","M"," R"},Left,Center];

(* Definition of f: not allowed producs are 0 *)

f[a[10],a[x_]]:=0/;x!=13;
f[a[13],a[x_]]:=0/;(x!=10 && x!=20 && x!=30);
f[a[12],a[x_]]:=0/;x!=11;
f[a[11],a[x_]]:=0/;(x!=12 && x!=23 && x!=32);
f[a[20],a[x_]]:=0/;(x!=12 && x!=23 && x!=32);
f[a[23],a[x_]]:=0/;(x!=10 && x!=20 && x!=30);
f[a[30],a[x_]]:=0/;(x!=33 && x!=43);
f[a[33],a[x_]]:=0/;(x!=10 && x!=20 && x!=30);
f[a[32],a[x_]]:=0/;(x!=31 && x!= 40);
f[a[31],a[x_]]:=0/;(x!=12 && x!=23 && x!=32);
f[a[40],a[x_]]:=0/;(x!=33 && x!=43);
f[a[43],a[x_]]:=0/;(x!=31 && x!=40);

f[f[x__,y_],z_]:=If[f[y,z]==0,0,f[x,y,z],f[x,y,z]];
f[x_,f[y_,z__]]:=If[f[x,y]==0,0,f[x,y,z],f[x,y,z]];
f[f[x__,y_],f[u_,v__]]:=If[f[y,u]==0,0,f[x,y,u,v],f[x,y,u,v]];
f[0,x__]:=0;f[x__,0]:=0;

(* Definition of inv *)

inv[f[x_,y___,z_]]:=If[f[z,x]==0,0,f[x,y,z],f[x,y,z]];
inv[0]:=0;
SetAttributes[inv,Listable];

(* Definitions of lists *)

l1={a[10],a[13],a[12],a[11],a[20],a[23],a[30],a[33],a[32],a[31],
a[40],a[43]};
cl1ini={a[10],a[20],a[30]};
cl2ini={a[13]};
cl3ini={a[33],a[43]};
cl1fin={a[13],a[23],a[33]};
cl2fin={a[10]};
cl3fin={a[30],a[40]};

(* Definitions of Looktrace and Lookdet *)

Looktrace[l_List,x_,y_,u_,v_]:=Block[{n,sust,lev,lt,ls},(
ls=Apply[Dot,l,1];
sust={a[10]->x,a[20]->y,a[30]->u,a[40]->v,a[13]->Adj[x],
a[23]->Adj[y],a[33]->Adj[u],a[43]->Adj[v],a[11]->Transpose[x],
a[31]->Transpose[u],a[12]->Conjugate[x],a[32]->Conjugate[u]};
lev=ls/.sust;
For[lt={};n=1,n<=Length[ls],n++,
If[Chop[trace[lev[[n]]]]!=Conjugate[Chop[trace[lev[[n]]]]],
AppendTo[lt,ls[[n]]] ] ];
lt)];

Lookdet[l_List,x_,y_,u_,v_,sgn_:-1]:=Block[{i,cm,sust,cev,lt},(
cm=Complement[Union[Flatten[Outer[Plus,Apply[Dot,autono[l],1],
sgn Apply[Dot,l,1] ] ] ],{0}];
sust={a[10]->x,a[20]->y,a[30]->u,a[40]->v,
a[13]->Adj[x],a[23]->Adj[y],a[33]->Adj[u],a[43]->Adj[v],
a[11]->Transpose[x],a[31]->Transpose[u],
a[12]->Conjugate[x],a[32]->Conjugate[u]};
cmev=cm/.sust;
For[lt={};i=1,i<=Length[cm],i++,
If[Chop[Det[cmev[[i]]]]!=Conjugate[Chop[Det[cmev[[i]]]]],
AppendTo[lt,cm[[i]]] ] ];lt)];

(* Definition of Newecs *)

Newecs[oldecs_,list_,x_,y_,u_,v_,vars_,complexvars_,ops___]:=
Block[{ecs2,ecs3,i,j,lts,lt,newec,end,what,detyn,trordet,vars2},(
detyn=(det/.{ops})/.Options[Newecs];
If[detyn,trordet[args__]:=Imdet[args],trordet[args__]:=Imtr[args],
trordet[args__]:=Imtr[args] ];
ecs2=oldecs;
ecs3={};
vars2=Join[vars,Re[complexvars],Im[complexvars]];
For[lts={};i=1,i<=Length[list],i++,(
 newec=trordet[list[[i]],x,y,u,v,complexvars];
 For[j=1;end=False;what=0,j<=Length[ecs2] && !end,j++,
  If[PolynomialQ[Factor[ecs2[[j]]/newec],vars2] && !(NumberQ[Factor
   [ecs2[[j]]/newec]]),what=j;end=True] ];
 If[end,ecs2[[what]]=newec;AppendTo[lts,list[[i]]] ] )];
For[lt={};i=1,i<=Length[list],i++,(
 newec=trordet[list[[i]],x,y,u,v,complexvars];
 For[j=1;end=False,j<=Length[ecs2] && !end,j++,
  If[PolynomialQ[Factor[newec/ecs2[[j]]],vars2],end=True] ];
 If[!end,AppendTo[ecs3,newec];AppendTo[lt,list[[i]]] ] )];
{ecs2,lts,ecs3,lt} )];

Options[Newecs]={det->False}

(* Definition of Eliminatesols *)

Eliminatesols[sols_]:=Block[{i,j,sols2,inter},(
sols2=Union[sols];
For[i=1,i<=Length[sols2],i++,If[Head[sols2[[i]]]==And,,
sols2[[i]]=And[sols2[[i]],sols2[[i]] ],
sols2[[i]]=And[sols2[[i]],sols2[[i]] ] ] ];
i=1;
While[i<Length[sols2],(
  j=i+1;
  While[j<=Length[sols2],(
    inter=Intersection[sols2[[i]],sols2[[j]] ];
    If[inter==Union[sols2[[i]]],sols2=Delete[sols2,j],,
      If[inter==Union[sols2[[j]]],
        sols2=Delete[sols2,i];j=i+1,,j++] ] )];
i++)];
For[i=1,i<=Length[sols2],i++,sols2[[i]]=Union[sols2[[i]]] ];
sols2 )];

(* Definitions of Red12 and Red21 *)

Red12[m0_]:=Block[{lu,m,m12,m21,m22,w2,w,delta,n},(
lu={};
m=m0;
m12:={{m[[1,2]],m[[1,3]]}};
m21:=Transpose[m12];
m22:={{m[[2,2]],m[[2,3]]},{m[[3,2]],m[[3,3]]}};
w2=DiagonalizeS[m22][[2]];
w=IncDim1[w2];
m=Chop[w.m.Transpose[w]];
If[Chop[m[[2,2]]-m[[3,3]]]==0,
  If[m12=={{0,0}},
    AppendTo[lu,2];AppendTo[lu,3],
  (* else *)
    If[m22=={{0,0},{0,0}},
      w2=DiagonalizeH[m21.Adj[m21]][[2]];
      w=IncDim1[w2];
      m=Chop[w.m.Transpose[w]],
    (* else *)
      If[m[[1,1]]!=0,
        delta=-Arg[m[[1,1]]]/2;
        w=DiagonalMatrix[{Exp[I delta],1,1}];
        m=w.m.Transpose[w],
      (* else *)
        n=(Chop[m12.Transpose[m12]])[[1,1]];
        If[n==0,
          AppendTo[lu,1],
        (* else *)
          delta=-Arg[n]/2;
          w=DiagonalMatrix[{Exp[I delta],1,1}];
          m=w.m.Transpose[w]
        ]
      ];
      If[Im[m12]=={{0,0}},
        w2=DiagonalizeH[Re[m21].Transpose[Re[m21]]][[2]];
        w=IncDim1[w2];
        m=Chop[w.m.Transpose[w]],
      (* else *)
        w2=DiagonalizeH[Im[m21].Transpose[Im[m21]]][[2]];
        w=IncDim1[w2];
        m=Chop[w.m.Transpose[w]]
      ]
    ]
  ] (* end m12=0 *)
(* else 1st if *) ];
m
)];

Red21[m_,me_]:=Block[{m2},
If[Chop[Abs[me[[1,1]]]-Abs[me[[2,2]]]]!=0,
  m,
(* else *)
  m2=Reverse[Transpose[Reverse[m]]];
  Reverse[Transpose[Reverse[Red12[m2]]]]
]];

(* Definition of LookCP *)

LookCP[m_]:=Block[{m2,dim,i,ecsdeg0,ecsdeg1,ecsdeg2,allecs,sol},(
m2=Chop[Arg[m]];
dim=Length[m];
For[i=1,i<=dim,i++,If[m[[i,i]]==0,m2[[i,i]]=delta[i] ] ];
ecsdeg0=Part[minors[m2],
  Flatten[Position[Apply[Plus,Thread[vec[minors[m2],dim]],1],0]]];
ecsdeg1=Part[minors[m2],
  Flatten[Position[Apply[Plus,Thread[vec[minors[m2],dim]],1],1]]];
ecsdeg2=Part[minors[m2],
  Flatten[Position[Apply[Plus,Thread[vec[minors[m2],dim]],1],2]]];
allecs=Join[ecsdeg1,ecsdeg2];
If[Complement[Chop[Sin[ecsdeg0]],{0}]=={},
If[allecs=={},True, (* there aren't equations *)
For[i=2;indeps={allecs[[1]]},i<=dim && i<=Length[allecs],i++,
  If[linealindep[Append[indeps,allecs[[i]] ],dim],
  AppendTo[indeps,allecs[[i]] ] ] ];
sol=Flatten[Solve[indeps==Table[0,{Length[indeps]}] ] ];
Complement[Chop[Chop[Sin[allecs/.sol]]],{0}]=={}],False] 
)]

End[]

EndPackage[]
\end{verbatim}

\end{document}